\documentclass[twocolumn,showpacs,preprintnumbers,amsmath,amssymb,superscriptaddress,floatfix]{revtex4}
\usepackage{graphicx}
\usepackage{dcolumn}
\usepackage{bm}
\usepackage{latexsym}
\def\Tr{\text{Tr}} 
\def\nn{\nonumber}

\newcommand{\Kanazawa}{\affiliation{Kanazawa University, Kanazawa 920-1192, Japan}}

\begin{document}

\title{A new scheme for color confinement and violation of the non-Abelian Bianchi identities}
\author{Tsuneo Suzuki}
\email[e-mail:]{suzuki04@staff.kanazawa-u.ac.jp}
\Kanazawa

\date{\today}

\begin{abstract} 
A new scheme for color confinement in QCD due to violation of the non-Abelian Bianchi identities is proposed.
The violation of the non-Abelian Bianchi identities (VNABI) $J_{\mu}$  is equal to Abelian-like monopole currents $k_{\mu}$ defined by the violation of the Abelian-like Bianchi identities.   Although VNABI is an adjoint operator satisfying the covariant conservation rule $D_{\mu}J_{\mu}=0$, it  gives us, at the same time, the Abelian-like conservation rule $\partial_{\mu}J_{\mu}=0$. The Abelian-like conservation rule $\partial_{\mu}J_{\mu}=0$ is also gauge-covariant.    There are $N^2-1$ conserved magnetic charges in the case of color $SU(N)$.  The charge of each component of VNABI is quantized \`{a} la Dirac. 
The color invariant eigenvalue $\lambda_\mu$ of VNABI also satisfies the Abelian conservation rule $\partial_\mu\lambda_\mu=0$ and the magnetic charge of the eigenvalue is also quantized \`{a} la Dirac. If the color invariant eigenvalue make condensation in the QCD vacuum, each color component of the non-Abelian electric field $E^a$ is squeezed by the corresponding color component of the sorenoidal current $J^a_{\mu}$. Then only the color singlets alone can survive as a physical state and non-Abelian color confinement is realized. This picture is completely new in comparison with the previously studied monopole confinement scenario based on an Abelian projection after some partial gauge-fixing, where Abelian neutral states can survive as physical. 

VNABI satisfying the Dirac quantization condition could be defined on lattice  as  lattice Abelian-like monopole currents without any gauge-fixing. Previous studies of the Abelian-like monopoles $k_{\mu}$ on lattice show that non-Abelian color confinement could be 
understood by the Abelian-like dual Meissner effect due to condensation of VNABI.  
\end{abstract}


\maketitle

\section{Introduction}
Color confinement in  quantum chromodynamics (QCD) 
is still an important unsolved  problem~\cite{CMI:2000mp}. 

As a picture of color confinement, 't~Hooft~\cite{tHooft:1975pu} and Mandelstam~\cite{Mandelstam:1974pi} conjectured that the QCD vacuum is a kind of a magnetic superconducting state caused by condensation of magnetic monopoles and  an effect dual to the Meissner effect works to confine color charges. 
However, in contrast to SUSY QCD
~\cite{Seiberg:1994rs} or Georgi-Glashow 
model~\cite{'tHooft:1974qc,Polyakov:1976fu} with scalar fields,
to find color magnetic monopoles which
condense is not straightforward in QCD. 

An interesting idea to realize this conjecture is to project QCD to the Abelian 
maximal torus group by a partial (but singular) gauge fixing~\cite{tHooft:1981ht}. 
In $SU(3)$ QCD, the maximal torus group is  Abelian $U(1)^2$. 
Then color magnetic monopoles appear as a topological object.
Condensation of the monopoles  causes  the dual Meissner 
effect~\cite{Ezawa:1982bf,Suzuki:1988yq,Maedan:1988yi}.

\par
Numerically, an Abelian projection in non-local 
gauges such as the maximally Abelian (MA)
gauge~\cite{Suzuki:1983cg,Kronfeld:1987ri,Kronfeld:1987vd} 
has been found to support the Abelian confinement scenario
beautifully~\cite{Suzuki:1992rw,Singh:1993jj,Chernodub:1997ay,Bali:1997cp,
Suzuki:1998hc,Koma:2003gq,Koma:2003hv}. The present author and his group have shown 
that the  Abelian dominance and the dual Meissner effect are observed  clearly  also in local unitary gauges such as $F12$ and Polyakov (PL) gauges~\cite{Sekido:2007mp}.  

\par
However, although numerically interesting, the idea of Abelian projection\cite{tHooft:1981ht} is theoretically very unsatisfactory. 1) In non-perturabative QCD, 
any gauge-fixing is  not necessary at all. There are infinite ways of such a partial gauge-fixing and whether the 't Hooft scheme is gauge independent or 
not is not known. 2)  After an Abelian projection, only one (in $SU(2)$) or two (in $SU(3)$) gluons are photon-like with respect to the residual $U(1)$ or $U(1)^2$ symmetry and the other gluons are massive charged matter fields. Such an asymmetry among octet gluons is  unnatural. 3) How one can construct Abelian monopole operators in terms of gluon fields is not clear at all.

The purpose of this work is to show a new scheme for color confinement due to the dual Meissner effect which is free from the above problems concerning the 'tHooft idea\cite{tHooft:1981ht},  

\vspace{0.5cm}

\section{Equivalence of  $J_{\mu}$ and $k_{\mu}$}
First of all, we prove that the Jacobi identities of covariant derivatives lead us to that  violation of the non-Abelian Bianchi identities (VNABI) $J_{\mu}$ is nothing but an Abelian-like monopole $k_{\mu}$ defined by violation of the Abelian-like Bianchi identities without gauge-fixing. 
 For simplicity we consider $SU(2)$ QCD. Define a covariant derivative operator $D_{\mu}=\partial_{\mu}-igA_{\mu}$. The Jacobi identities are expressed as 
\begin{eqnarray}
\epsilon_{\mu\nu\rho\sigma}[D_{\nu},[D_{\rho},D_{\sigma}]]=0. \label{eq-Jacobi}
\end{eqnarray}
By direct calculations, one gets
\begin{eqnarray*}
[D_{\rho},D_{\sigma}]&=&[\partial_{\rho}-igA_{\rho},\partial_{\sigma}-igA_{\sigma}]\\
&=&-ig(\partial_{\rho}A_{\sigma}-\partial_{\sigma}A_{\rho}-ig[A_{\rho},A_{\sigma}])+[\partial_{\rho},\partial_{\sigma}]\\
&=&-igG_{\rho\sigma}+[\partial_{\rho},\partial_{\sigma}],
\end{eqnarray*}
where the second commutator term of the partial derivative operators can not be discarded, since gauge fields may contain a line singularity. Actually, it is the origin of the violation of the non-Abelian Bianchi identities (VNABI) as shown in the following.
The relation $[D_{\nu},G_{\rho\sigma}]=D_{\nu}G_{\rho\sigma}$ and the Jacobi identities (\ref{eq-Jacobi}) lead us to
\begin{eqnarray}
D_{\nu}G^{*}_{\mu\nu}&=&\frac{1}{2}\epsilon_{\mu\nu\rho\sigma}D_{\nu}G_{\rho\sigma} \nn\\
&=&-\frac{i}{2g}\epsilon_{\mu\nu\rho\sigma}[D_{\nu},[\partial_{\rho},\partial_{\sigma}]]\nn\\
&=&\frac{1}{2}\epsilon_{\mu\nu\rho\sigma}[\partial_{\rho},\partial_{\sigma}]A_{\nu}\nn\\
&=&\partial_{\nu}f^{*}_{\mu\nu}, \label{eq-JK}
\end{eqnarray}
where $f_{\mu\nu}$ is defined as $f_{\mu\nu}=\partial_{\mu}A_{\nu}-\partial_{\nu}A_{\mu}=(\partial_{\mu}A^a_{\nu}-\partial_{\nu}A^a_{\mu})\sigma^a/2$. The left-hand side of Eq.(\ref{eq-JK}) if vanishing is the non-Abelian Bianchi identities, whereas the right-hand side of Eq.(\ref{eq-JK}) if vanishing is the Abelian-like Bianchi identities.
Namely Eq.(\ref{eq-JK}) shows that the violation of the non-Abelian Bianchi identities is equivalent to that of the Abelian-like Bianchi identities.

Denote the violation of the non-Abelian Bianchi identities as  $J_{\mu}$:
\begin{eqnarray}
J_{\mu} &=& \frac{1}{2}J_{\mu}^a\sigma^a
=D_{\nu}G^*_{\mu \nu}. \label{nabi}
\end{eqnarray}
Eq.(\ref{nabi}) is gauge covariant and therefore a non-zero  $J_{\mu}$ is a gauge-invariant property. An Abelian-like monopole $k_{\mu}$ without any gauge-fixing is defined as the violation of the Abelian-like Bianchi identities:
\begin{eqnarray}
k_{\mu}=\frac{1}{2}k_{\mu}^a\sigma^a&=& \partial_{\nu}f^*_{\mu\nu}
=\frac{1}{2}\epsilon_{\mu\nu\rho\sigma}\partial_{\nu}f_{\rho\sigma}. \label{ab-mon}
\end{eqnarray}
Eq.(\ref{eq-JK}) shows that
\begin{eqnarray}
J_{\mu}=k_{\mu}. \label{JK}
\end{eqnarray}
 
Several comments are in order.
\begin{enumerate}
\item Eq.(\ref{JK}) can be considered as an extension of the important relation derived recently by Bonati et al.\cite{Bonati:2010tz} in the framework of an Abelian projection to a simple case without any Abelian projection.
Actually it is possible to prove directly without the help of the Jacobi identities
\begin{eqnarray*}
J_{\mu}^a-k_{\mu}^a&=& \Tr\sigma^a D_{\nu}G^{*}_{\mu\nu}-\partial_{\nu}f^{*a}_{\mu\nu} \\
&=&-ig\Tr\sigma^a[A_{\nu}, G^{*}_{\mu\nu}]\\
&&-ig\epsilon_{\mu\nu\rho\sigma}\Tr\sigma^a[\partial_{\nu}A_{\rho}, A_{\sigma}]\\
&=&0. 
\end{eqnarray*}
Abelian monopoles in any Abelian projection such as MA gauge are  related to VNABI as shown in Ref.\cite{Bonati:2010tz}.  Hence VNABI itself is  expected to be a key quantity in color confinement.
\item
VNABI $J_{\mu}$ transforms as an adjoint operator, so that does the Abelian-like monopole current $k_{\mu}$. This can be proved also directly. Consider a regular gauge transformation
\begin{eqnarray*}
A'_{\mu}&=&VA_{\mu}V^{\dag}-\frac{i}{g}\partial_{\mu}VV^{\dag}.
\end{eqnarray*}
Then
\begin{eqnarray}
k'_{\mu}&=&\epsilon_{\mu\nu\rho\sigma}\partial_{\nu}\partial_{\rho}A'_{\sigma}\nn\\
&=&\epsilon_{\mu\nu\rho\sigma}\partial_{\nu}\partial_{\rho}(VA_{\mu}V^{\dag}-\frac{i}{g}\partial_{\mu}VV^{\dag})\nn\\
&=&V(\epsilon_{\mu\nu\rho\sigma}\partial_{\nu}\partial_{\rho}A_{\sigma})V^{\dag}\nn\\
&=&Vk_{\mu}V^{\dag}.\label{vkv}
\end{eqnarray}

\item
The above equivalence shows VNABI is essentially Abelian-like. It was already argued that singularities of gauge fields corresponding to VNABI must be Abelian\cite{DiGiacomo:2008wh}, although the reasoning is different.
\item The covariant conservation rule $D_{\mu}J_{\mu}=0$ is proved as follows\cite{Bonati:2010tz}:
\begin{eqnarray}
D_{\mu}J_{\mu}&=&D_{\mu}D_{\nu}G^*_{\nu\mu}
=\frac{ig}{2}[G_{\nu\mu},G^*_{\nu\mu}]\nn\\
&=&\frac{ig}{4}\epsilon_{\nu\mu\rho\sigma}[G_{\nu\mu},G_{\rho\sigma}]
=0,  \label{NA-cons}
\end{eqnarray}
where 
\begin{eqnarray}
\partial_{\mu}\partial_{\nu}G^{*}_{\mu\nu}=0 \label{PNA}
\end{eqnarray}
 is used.
The Abelian-like monopole  satisfies the Abelian-like  conservation rule
\begin{eqnarray}
\partial_{\mu}k_{\mu}=\partial_{\mu}\partial_{\nu}f^{*}_{\mu\nu}=0\label{PAA}
\end{eqnarray}
due to the antisymmetric property of the Abelian-like field strength\cite{Arafune:1974uy}. Hence VNABI satisfies also the same Abelian-like conservation rule 
\begin{eqnarray}
\partial_{\mu}J_{\mu}=0. \label{A-cons}
\end{eqnarray}
Both Eqs.(\ref{NA-cons}) and (\ref{A-cons}) are compatible, since
the difference between both quantities
\begin{eqnarray}
[A_{\mu}, J_{\mu}]&=&\frac{1}{2}\epsilon_{\mu\nu\rho\sigma}[A_{\mu},\partial_{\nu}f_{\rho\sigma}]\nn\\
&=&\epsilon_{\mu\nu\rho\sigma}[A_{\mu}, \partial_{\nu}\partial_{\rho}A_{\sigma}]\nn\\
&=&-\frac{1}{2}\epsilon_{\mu\nu\rho\sigma}\partial_{\nu}\partial_{\mu}[A_{\rho},A_{\sigma}]\nn\\
&=&\frac{i}{g}(\partial_{\mu}\partial_{\nu}G^*_{\mu\nu}-\partial_{\mu}\partial_{\nu}f^*_{\mu\nu})\nn\\
&=& 0\nn,
\end{eqnarray}
where (\ref{PNA}) and (\ref{PAA}) are used.
 Hence the Abelian-like conservation relation (\ref{A-cons}) is also gauge-covariant.

\item The Abelian-like conservation relation (\ref{A-cons}) gives us three conserved magnetic charges in the case of color $SU(2)$ and $N^2-1$ charges in the case of color $SU(N)$. But these are kinematical  relations coming from the derivative with respect to the divergence of an antisymmetric tensor~\cite{Arafune:1974uy}.  The number of conserved charges is different from that of the Abelian projection scenario~\cite{tHooft:1981ht}, where only $N-1$ conserved charges exist in the case of color $SU(N)$.
\end{enumerate}

\section{Dirac quantization condition}
Next we show that the magnetic charges derived from $k_4=J_4$ satisfy the Dirac quantization condition with respect to  magnetic and electric charges. Consider a space-time point $O$ where the Bianchi identities are violated and a three-dimensional sphere $V$ of a large radius $r$ from $O$. Since $k_4=J_4$ is given by the total derivative, the  behavior of the gauge field at the surface of the sphere  is relevant. When $r\to\infty$, the non-Abelian field strength should 
vanish since otherwise the action diverges. 
Then the magnetic charge could be evaluated by a gauge field described by a pure gauge $A_{\mu}=\Omega\partial_\mu\Omega^{\dag}/ig$, where $\Omega$ is a gauge transformation matrix satisfying $\Omega[\partial_\mu, \partial_\nu]\Omega^{\dag}= 0$ at $r\to\infty$. Then the magnetic charge $g^d_m$ in a color direction is evaluated as follows:
\begin{eqnarray}
g^d_m&=&\int_V d^3x k^d_4=\int d^3x \frac{1}{2}\epsilon_{4\nu\rho\sigma}\partial_{\nu}(\partial_\rho A^d_\sigma-\partial_\sigma A^d_\rho) \nn\\
&=&\int_V d^3x \frac{1}{2ig}\epsilon_{ijk}\partial_{i} \Tr \sigma^d(\partial_j\Omega\partial_k\Omega^{\dag}-
\partial_k\Omega\partial_j\Omega^{\dag} \nn\\
&& + \Omega[\partial_j,\partial_k]\Omega^{\dag}) \nn\\
&=&\int_V d^3x \frac{1}{2g}\epsilon_{ijk}\{\epsilon^{abc}\partial_i(\hat{\phi}^a\partial_{j}\hat{\phi}^b\partial_{k}\hat{\phi}^c\nn\\ 
&&+ \partial_{i} \Tr \sigma^d\Omega[\partial_j,\partial_k]\Omega^{\dag})\} \nn\\
&=&\int_{\partial V}d^2S \frac{1}{2g}\epsilon_{ijk}\epsilon^{abc}\hat{\phi}^a\partial_{j}\hat{\phi}^b\partial_{k}\hat{\phi}^c, \label{GM}
\end{eqnarray}
where 
$\Omega[\partial_j, \partial_k]\Omega^{\dag}= 0$ on the surface at $r\to\infty$ is used and 
$\hat{\phi}$ is a Higgs-like field defined as 
 \begin{eqnarray*}
\hat{\phi}&=&\hat{\phi}^i\sigma^i\\
&=&\Omega\sigma^d\Omega^{\dag}.
\end{eqnarray*}
$\hat{\phi}^2=1$ is shown easily. Since the field $\hat{\phi}$ is a single-valued function, Eq.(\ref{GM}) is given by the wrapping number $n$ characterizing the homotopy class
of the mapping between the spheres described by $\hat{\phi}^2=(\hat{\phi}^1)^2+(\hat{\phi}^2)^2+(\hat{\phi}^3)^2=1$ and $\partial V=S^2$: $\pi_2(S^2)=Z$. Namely 
\begin{eqnarray}
g^d_mg=4\pi n. \label{g-gm}
\end{eqnarray}
This is just the Dirac quantization condition.
Note that the minimal color electric charge in any color direction is $g/2$. Hence the kinematical conservation rule  is also topological.

\vspace{.5cm}

What happens in the case of color $SU(3)$? Then it is easy to prove that the three $SU(2)$ subspaces (isospin, U-spin, V-spin) play the role in the above mapping  and all eight magnetic charges are quantized similarly \`{a} la Dirac.
In  $SU(3)$,  a gauge field for $r\to\infty$ becomes  $A_{\mu}=\Omega\partial_\mu\Omega^{\dag}/ig$, where now $\Omega$ is a $3\times 3$ gauge transformation matrix of $SU(3)$. Then for example, consider a magnetic charge in the $\lambda^1$ direction, where $\lambda^a$ is the GellMann matrix. Then
define  a Higgs-like field $\hat{\phi}$ as 
 \begin{eqnarray}
\hat{\phi}&=&\hat{\phi}^i\lambda^i\\
&=&\Omega\lambda^1\Omega^{\dag}.\label{su3phi}
\end{eqnarray}
The magnetic charge $g^1_m$ in the $\lambda^1$ color direction is evaluated as follows:
\begin{eqnarray}
g^1_m&=&\int_V d^3x k^1_4=\int d^3x \frac{1}{2}\epsilon_{4\nu\rho\sigma}\partial_{\nu}(\partial_\rho A^1_\sigma-\partial_\sigma A^1_\rho) \nn\\
&=&\int_V d^3x \frac{1}{2ig}\epsilon_{ijk}\partial_{i} \Tr \lambda^1(\partial_j\Omega\partial_k\Omega^{\dag}-
\partial_k\Omega\partial_j\Omega^{\dag} \nn\\
&& + \Omega[\partial_j,\partial_k]\Omega^{\dag}) \nn\\
&=&\int_V d^3x \frac{1}{2g}\epsilon_{ijk}\{\epsilon^{abc}\partial_i(\hat{\phi}^a\partial_{j}\hat{\phi}^b\partial_{k}\hat{\phi}^c\nn\\ 
&&+ \partial_{i} \Tr \lambda^1\Omega[\partial_j,\partial_k]\Omega^{\dag})\} \nn\\
&=&\int_{\partial V}d^2S \frac{1}{2g}\epsilon_{ijk}\epsilon^{abc}\hat{\phi}^a\partial_{j}\hat{\phi}^b\partial_{k}\hat{\phi}^c.
\end{eqnarray}
Now one sees from (\ref{su3phi})
\begin{eqnarray}
\hat{\phi}^2=\left(
  \begin{array}{ccc}
     1  &   0 & 0   \\
     0  &  1  & 0   \\
     0  &   0 &  0  \\
  \end{array}
\right). \label{phi3}
\end{eqnarray}
The condition (\ref{phi3}) gives us  
\begin{eqnarray*}
(\hat{\phi}^1)^2+(\hat{\phi}^2)^2+(\hat{\phi}^3)^2&=& 1,\\
\hat{\phi}^4=\hat{\phi}^5=\hat{\phi}^6=\hat{\phi}^7=\hat{\phi}^8&=& 0.
\end{eqnarray*}
Namely the subspace composed of $(\hat{\phi}^1, \hat{\phi}^2,  \hat{\phi}^3)$ is a sphere $S^2$ and the mapping is just like that in the case of color $SU(2)$.
Hence $g_m^1$ satifies the Dirac quantization condition 
\begin{eqnarray}
g^1_mg=4\pi n. 
\end{eqnarray}
The same condition holds good for all other magnetic charges.

\vspace{.5cm}
\section{The confining vacuum proposed}
Now we propse a new mechanism of color confinement in which VNABI $J_{\mu}$ play an important role in the vacuum. However  $J_{\mu}$ has a color electric charge as well as the magnetic charge. Hence the  condensation of $J_{\mu}$ 
themselves  seems to give us at the same time a spontaneous breaking of $SU(2)$ color electric symmetry. It is contradictory to the usual expectation concerning the color confinement, where the electric color symmetry is kept exact. 

   Since VNABI transforms as an adjoint operator,   it can be diagonalized by a unitary matrix $V_d(x)$ as follows:
\begin{eqnarray*}
V_d(x)J_{\mu}(x)V_d^{\dag}(x)=\lambda_{\mu}(x)\frac{\sigma_3}{2},
\end{eqnarray*}
where $\lambda_{\mu}(x)$ is the eigenvalue of $J_{\mu}(x)$ and is then color invariant but magnetically charged. Note that $V_d(x)$ does not depend on $\mu$ due to the Coleman-Mandula theorem\cite{Coleman}
\footnote{Applying  the Coleman-Mandula theorem  to QCD, I note that Kugo-Ojima\cite{Kugo} showed a manifestly covariant and local canonical operator formalism of non-Abelian gauge theories. Although introducing an indefinite metric is inevitable, the unitarity of the physical $S$-matrix is proved.  Moreover the string-like behavior existing in  gauge fields in the case of color confinement is not rejected in the framework.}. 
Then one gets
\begin{eqnarray}
\Phi(x)&\equiv& V_d^{\dag}(x)\sigma_3 V_d(x) \label{Phi}\\
J_{\mu}(x)&=&\frac{1}{2}\lambda_{\mu}(x)\Phi(x). \label{lambda}
\end{eqnarray}
Namely the color electrically charged part and the magnetically charged part are separated out. From (\ref{lambda}) and (\ref{A-cons}), one gets
\begin{eqnarray}
\partial_{\mu}J_{\mu}(x)&=&\frac{1}{2}(\partial_{\mu}\lambda_{\mu}(x)\Phi(x) + \lambda_{\mu}(x)\partial_{\mu}\Phi(x))\nn\\
&=& 0.
\end{eqnarray}
Since $\Phi(x)^2=1$, 
\begin{eqnarray*}
\partial_{\mu}\lambda_{\mu}(x)&=&-\lambda_{\mu}(x)\Phi(x)\partial_{\mu}\Phi(x)\\
&=&0.
\end{eqnarray*}
Hence the eigenvalue $\lambda_{\mu}$ itself satisfies the Abelian conservation
 rule. Moreover when use is made of (\ref{vkv}) and (\ref{GM}), it is easy to prove that 
\begin{eqnarray}
\lambda_{\mu}(x)\frac{\sigma_3}{2}&=&\epsilon_{\mu\nu\rho\sigma}\partial_{\nu}\partial_{\rho}A'_{\sigma}(x),\label{lambda3}
\end{eqnarray}
where $A'_{\mu}=V_dA_{\mu}V^{\dag}_d-\frac{i}{g}\partial_{\mu}V_dV_d^{\dag}$.
Here Eq.(\ref{lambda3}) means the singularity appears only in the diagonal component of the gauge field $A'_{\mu}$. This may be related to the work\cite{DiGiacomo:2008wh}.
Hence if one considers for large $r$
 \begin{eqnarray*}
A'_{\mu}&\to&\Omega\partial_\mu\Omega^{\dag}/ig,\\ 
\hat{\phi}&=&\hat{\phi}^i\sigma^i
=\Omega\sigma^3\Omega^{\dag},
\end{eqnarray*}
one can easily see from (\ref{GM}) that the magnetic charge from the eigenvalue $\lambda_{\mu}$ also satisfies the Dirac quantization condition  (\ref{g-gm}).
The condensation of the color-singlet magnetic currents $\lambda_{\mu}$ does not give rise to a spontaneous breaking of the color electric symmetry. 
Condensation of the color invariant  magnetic currents $\lambda_{\mu}$ may be a key mechanism of  the physical confining vacuum. This is a new scheme of color confinement we propose.
To clarify clearly the difference of this scheme from the previous 'tHooft Abelian projection with some partial gauge-fixing, we show  Table\ref{comp} in which typical different points are written..

\vspace{.5cm}
\begin{table*}
\caption{\label{comp}Comparison between the 'tHooft Abelian projection studies and the present work in $SU(2)$ QCD.  $\hat{\phi}'=V_p^{\dag}\sigma_3 V_p$, where $V_p$ is a partial gauge-fixing matrix of an Abelian projection. $(u_c, d_c)$ is a color-doublet quark pair. MA means maximally Abelian. }
\begin{ruledtabular}
\begin{tabular}{|c|c|c|c|}
  &\multicolumn{2}{c|}{The 'tHooft Abelian projection scheme}  &  This work and Refs.\cite{Suzuki:2007jp,Suzuki:2009xy} \ \   \\
  &Previous works\cite{Suzuki:1983cg,Kronfeld:1987ri,Kronfeld:1987vd,Suzuki:1992rw,Singh:1993jj,Chernodub:1997ay,Bali:1997cp,Suzuki:1998hc,Koma:2003gq,Koma:2003hv,Sekido:2007mp} & Reference \cite{Bonati:2010tz} & \\
  \hline
  Origin of  $k_{\mu}$ & A singular gauge transformation& $k_{\mu}=\Tr J_{\mu}\hat{\phi}'$ & $k_{\mu}^a=J_{\mu}^a$\\
  \hline
No. of conserved  $k_{\mu}$   & \multicolumn{2}{c|}{$1$}   &  $3$  \\ 
Role of $A^a_{\mu}$ &\multicolumn{2}{c|}{One photon $A_{\mu}$ with $k_{\mu}$ $+$ 2 massive $A^{\pm}_{\mu}$} &  Three gluons $A_{\mu}^a$ with $k_{\mu}^a$ \\
Flux squeezing&\multicolumn{2}{c|}{One electric field $E_{\mu}$ }& Three electric fields $E^a_{\mu}$\\
   \hline
 Number of physical mesons &\multicolumn{2}{c|}{ 2 Abelian neutrals, $\bar{u}_cu_c$ and $\bar{d}_cd_c$}& 1 color singlet $\bar{u}_cu_c+\bar{d}_cd_c$\\
   \hline
 Expected confining vacuum &\multicolumn{2}{c|}{Condensation of Abelian monopoles }& Condensation of color-invariant $\lambda_{\mu}$\\
 \hline 
Privileged gauge choice   &  No special one &  MA gauge  & No gauge-fixing   \\
\end{tabular}
\end{ruledtabular} 
\end{table*}

\section{VNABI and Abelian monopoles in various Abelian projection scheme}
Let us discuss here the relation  derived by Bonati et al.\cite{Bonati:2010tz}:
\begin{eqnarray}
k^{AB}_{\mu}(x)&=&\Tr\{J_{\mu}(x)\Phi^{AB}(x)\}, \label{kab}
\end{eqnarray}
where $k^{AB}_{\mu}(x)$ is an Abelian monopole, $\Phi^{AB}(x)=V^{\dag}_{AB}(x)\sigma_3V_{AB}(x)$ and $V_{AB}(x)$ is a partial gauge-fixing matrix in some Abelian projection  like the MA gauge. Making use of Eq.(\ref{lambda}), we get
\begin{eqnarray}
k^{AB}_{\mu}(x)&=&\lambda_{\mu}(x)\tilde{\Phi}^3(x), 
\end{eqnarray}
where 
\begin{eqnarray*}
\tilde{\Phi}(x)&=&V_{AB}(x)V_d^{\dag}(x)\sigma_3V^{\dag}_{AB}(x)V_d(x)\\
&=&\tilde{\Phi}^a(x)\sigma^a.
\end{eqnarray*}
The relation (\ref{kab}) is very important, since existence of an Abelian monopole in any Abelian projection scheme is guaranteed by that of VNABI $J_{\mu}$ in the continuum limit.
Hence if in any special gauge such as MA gauge, Abelian monopoles remain non-vanishing in the continuum as suggested by many numerical data~\cite{Suzuki:1992rw,Singh:1993jj,Chernodub:1997ay,Bali:1997cp,
Suzuki:1998hc,Koma:2003gq,Koma:2003hv}, VNABI also remain non-vanishing in the continuum.

\vspace{.5cm}

\section{Lattice numerical results}
To prove the correctness of the new confinement scheme, we have made numerical simulations of lattice QCD.
\subsection{Definition of VNABI on lattice}
There is a long history studying VNABI on lattice. Skala et al. \cite{Skala:1996ar} adopted a naive definition of VNABI using a three dimensional cube composed of six plaquettes and found that VNABI strongly influence the confining property.
Gubarev and Morozov\cite{Gubarev:2005it} adopted a more sofisticated lattice definition of VNABI and also found that the suppression of VNABI is likely to destroy confinement in $D=4$ dimensions.

Since VNABI is found to be equivalent to Abelian-like monopoles without gauge-fixing, it is possible to define VNABI on lattice following the method of defining lattice Abelian-like monopoles\cite{DeGrand:1980eq,Suzuki:2007jp,Suzuki:2009xy}.

Write an $SU(2)$ link field as
\begin{eqnarray*}
U(s,\mu)&=&U^0(s,\mu)+i\vec{\sigma}\cdot\vec{U}(s,\mu).
\end{eqnarray*}
Then Abelian-like link fields are defined\cite{Suzuki:2007jp,Suzuki:2009xy} as
\begin{eqnarray}
\theta_{\mu}^a(s)&=&\arctan(\frac{U^a(s,\mu)}{U^0(s,\mu)}).
\end{eqnarray}
Abelian-like plaquette variables 
\begin{eqnarray*}
\theta^a_{\mu\nu}(s)&=&\partial_{\mu}\theta^a_{\nu}(s)-\partial_{\nu}\theta^a_{\mu}(s)
\end{eqnarray*}
are decomposed into
\begin{eqnarray}
\theta^a_{\mu\nu}(s)&=&\bar{\theta}^a_{\mu\nu}(s)+2\pi 
n^a_{\mu\nu}(s)\ \ (|\bar{\theta}^a_{\mu\nu}|<\pi),
\end{eqnarray}
where $n^a_{\mu\nu}(s)$ is an integer 
corresponding to the number of the Dirac string.
Abelian-like monopoles without gauge-fixing are defined\cite{DeGrand:1980eq} as
\begin{eqnarray}
k^a_{\mu}(s)= (1/2)\epsilon_{\mu\alpha\beta\gamma}\partial_{\alpha}
\bar{\theta}^a_{\beta\gamma}(s+\hat\mu).
\end{eqnarray}
Hence  VNABI is defined on lattice similarly as
\begin{eqnarray}
J_{\mu}(s)&= &(1/2)\epsilon_{\mu\alpha\beta\gamma}\partial_{\alpha}
\bar{\theta}_{\beta\gamma}(s+\hat\mu). \label{lat-J}
\end{eqnarray}
This quantity transforms as an adjoint operator as shown in the continuum in Eq.(\ref{vkv}), neglecting higher order terms with respect to the lattice distance.
It is to be noted that
a color component $J^a_{\mu}(s)$ of the lattice VNABI is integer. This corresponds to the Dirac quantization codition between electric and magnetic charges Eq.({\ref{g-gm})\cite{DeGrand:1980eq}. Eq.(\ref{lat-J}) leads us to the Abelian-like conservation rule $\partial'_{\mu}J_{\mu}(s)= 0$
where $\partial'_{\mu}$ denotes the lattice backward difference.
\subsection{Existence of VNABI in the continuum limit}

The present author with collaborators  have made high presision lattice simulations of Abelian-like monopoles without gauge-fixing~\cite{Suzuki:2007jp,Suzuki:2009xy}.  The results obtained are very interesting, although the authors of Ref.~\cite{Suzuki:2007jp,Suzuki:2009xy} could not understand at that time why gauge-variant monopoles are important. Let me review some of them explicitly in short, since the same results are obtained with respect to VNABI.
\begin{enumerate}
\item A static quark-antiquark potential derived from  Abelian-like Polyakov loop correlators 
gives us the same string tension as the non-Abelian one within the statistical error bar.  
\item The Hodge decomposition of the Abelian-like Polyakov loop
correlator to the regular photon and the singular
monopole parts also reveals that only the monopole
part is responsible for the string tension. 
\item The scaling behaviors of the string tensions are shown rather beautifully and the volume dependence is not seen, although more detailed studies are necessary.
\item 
The flux-tube profile is studied by
evaluating connected correlation
functions~\cite{Cea:1995zt, DiGiacomo:1989yp}.
Abelian-like electric fields defined in
an arbitrary color direction are squeezed by monopole
supercurrents with the same color direction.  
Only the electric field parallel to the static quark pair $E^a_{Az}$ is found to be squeezed. 
The penetration length
 $\lambda$ is determined as $\lambda=0.128(2)$~[fm]. This 
is similar to those obtained in various Abelian projection 
cases~\cite{Sekido:2007mp}.
\item To see what squeezes the Abelian-like electric field, 
Ref.\cite{Suzuki:2007jp,Suzuki:2009xy} studied the Abelian-like (dual) Amp\`ere law 
\begin{eqnarray*}
\vec{\nabla}\times\vec{E}^a_{A}=
\partial_{4}\vec{B}^a_{A}+2\pi\vec{k}^a\;, 
\end{eqnarray*}
where $B^a_{Ai}(s)=(1/2)\epsilon_{ijk}\bar{\theta}^a_{jk}(s)$.
Each term is evaluated on the same mid-plane
as for the electric field.
It is found that only the azimuthal components are non-vanishing,
which are plotted in Fig.~\ref{fig-1}~\cite{Suzuki:2009xy}.
The curl of the electric field
is non-vanishing and is reproduced only by monopole currents.  
These behaviors are clearly a signal of the Abelian-like dual
Meissner effect, which are quite the same as those 
observed in the MA gauge~\cite{Koma:2003gq,Koma:2003hv}. 

\begin{figure}[t]
\includegraphics[height=4.cm]{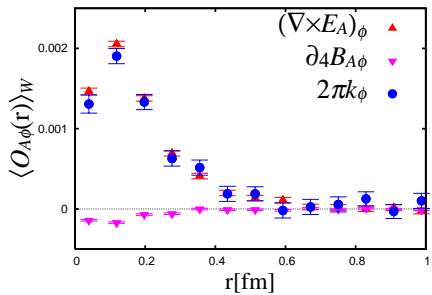}
\caption{\label{fig-1} The curl of the Abelian-like electric field, 
magnetic displacement currents and monopole currents 
for $W(R=5a,T=5a)$.}
 \vspace{-0.3cm}
\end{figure}

\item The coherence length~$\xi$ is evaluated from the correlation function between
the Wilson loop and the squared monopole density 
$(k^a_{\mu})^2(s)$
~\cite{Chernodub:2005gz}.
The  coherence length extracted from the correlation function $g(r)=c_{1}'\exp (-\sqrt{2}r/\xi) + c_{0}'$
is $\xi/\sqrt{2}=0.102(3)$~[fm].  
The GL parameter $\sqrt{2}\kappa=\lambda/\xi=1.25(6)$ is 
close to the values obtained with gauge fixing\cite{Sekido:2007mp}.
Since the Wilson loop used here may still be small, 
what one can say is that the vacuum type is near 
the border between the type~1 and~2.

\item Since all color components of Abelian-like monopoles make condensation, only a state without any color electric field can survive in physical world. They are only non-Abelian color singlets. Hence non-Abelian color confinement can be understood by the Abelian-like dual Meissner effect due to 
condensation of Abelian-like monopoles and VNABI. 
\end{enumerate} 

These numerical results, although the volume effects and the continuum limit are not yet completely studied, suggest that (1) the lattice definition\cite{DeGrand:1980eq} works good also in the case of VNABI or Abelian-like monopoles without any gauge-fixing
and (2) the dual Meissner effect caused by condensation of VNABI is a key mechanism of color confinement of QCD.


 \vspace{.5cm}

Some comments are in order.
\begin{itemize}
\item
To check the dual Meissner effect in the continuum limit, it is interesting to derive  an infrared effective theory in terms of VNABI.  Such a study was done in the case of MA gauge~\cite{Shiba:1994db,Chernodub:2000ax} with the help of an inverse Monte-Carlo method
and the block-spin transformation. However in the case of VNABI, it transforms as an adjoint operator and the effective action for VNABI should satisfy the color symmetry. 
\item 
Since VNABI transforms as an adjoint operator,  a quantity $J_{\mu}^2$ is  gauge-invariant and physical. Taking a sum with respect to the Lorentz index, one gets a gauge-invariant and a Lorentz scalar observable $\sum_{\mu} J_{\mu}^2$. It is interesting to study the quantity in  numerical simulations, since it may correspond to a monopole scalar field expected theoretically\cite{Suzuki:1988yq,Maedan:1988yi}. 
\item Also interesting are  numerical studies  like those done in Refs.\cite{Skala:1996ar, Gubarev:2005it} analyzing the effect of VNABI, although their lattice definitions of VNABI are different.
\end{itemize}
\vspace{.5cm}

\section{Conclusion}
Finally let me summarize new findings of this note.
\begin{enumerate}
  \item VNABI is equal to the Abelian-like monopole coming from the violation of the Abelian-like Bianchi identities.
  \item VNABI satisfies the Abelian-like conservation rule as well as the covariant one. Hence there are $N^2-1$ conserved magnetic charges in the case of color $SU(N)$. 
  \item   All magnetic charges  are quantized \`{a} la Dirac.
  \item VNABI can be defined on lattice as lattice Abelian-like monopoles. Previous numerical results suggest that the dual Meissner effect due to condensation of VNABI must be the color confinement mechanism of QCD. The role of Abelian monopoles is played by VNABI. This must be a new scheme for color confinement in QCD. 
  \item Condensation of the color invariant  magnetic currents $\lambda_{\mu}$ which are the eigenvalue of VNABI $J_{\mu}$ 
may be a key mechanism of  the physical confining vacuum.
\end{enumerate}

\begin{acknowledgments}
The author would like to thank Dr.Yoshiaki Koma for discussions on the Dirac string, Dr. Valentine Zakharov for suggesting a problem concerning the condensation of non-Abelian monopoles and Dr. Seiko Kato for giving information of a new definition of lattice monopoles.
\end{acknowledgments}

\end{document}